\documentclass[english,aps,prl,10pt,tightenlines,twocolumn,superscriptaddress,floatfix,fleqn]{revtex4-2}
\pdfoutput=1
\usepackage[utf8]{inputenc}
\usepackage[T1]{fontenc}
\usepackage{amsmath}
\usepackage{amssymb}
\usepackage{amsfonts}
\usepackage{bm}
\usepackage{bbm}
\usepackage{epsfig}
\usepackage{grffile}
\usepackage{times}

\usepackage[usenames,dvipsnames]{color}
\definecolor{dblue}{rgb}{0,0.1,.6}
\definecolor{dred}{rgb}{.6,0.1,0}

\usepackage[colorlinks=true,citecolor=dblue,linkcolor=dblue,urlcolor=dblue]{hyperref}
\usepackage[all]{hypcap}

\newcommand{\ud}{\mathrm{d}}
\newcommand{\mc}[1]{\mathcal{#1}}
\newcommand{\dist}{\operatorname{dist}}

\renewcommand{\vec}[1]{{\boldsymbol{#1}}}

\newcommand{\C} {\mc{C}}
\renewcommand{\O} {\mc{O}}
\newcommand{\W} {\mc{W}}
\newcommand{\vg}{\vec{g}}
\newcommand{\vs}{\vec{\sigma}}
\newcommand{\vx}{\vec{x}}

\newcommand{\const}{\textnormal{const}}

\newcommand{\tI}{\textnormal{I}}

\newcommand{\Emph}[1]{\emph{\textbf{#1}}}

\renewcommand{\section}[1]{\Emph{#1.}~--}

\newcommand{\qlab}  {National Quantum Laboratory, University of Maryland, College Park, MD 20742, USA}
\newcommand{\umd}   {Department of Physics, University of Maryland, College Park, MD 20742, USA}
\newcommand{\duke}  {Department of Physics, Duke University, Durham, North Carolina 27708, USA}
\newcommand{\dukeSoc}{Department of Sociology, Duke University, Durham, North Carolina 27708, USA}

\begin{document}

\title{Rare-event sampling for stochastic dynamics in network systems using cluster updates}
\author{Jiazheng Sun}
\affiliation{\duke}
\author{James Moody}
\affiliation{\dukeSoc}
\author{Thomas Barthel}
\affiliation{\duke}
\affiliation{\umd}
\affiliation{\qlab}
\date{August 16, 2026}

\begin{abstract}
Understanding the stochastic evolution in complex networks is a central challenge across physics, biology, engineering, social science, and finance. The most consequential macroscopic events, like cascading failures in communication networks, widespread epidemic outbreaks, and rapid shifts in societal opinions, often emerge from a confluence of rare, localized stochastic processes and need to pass certain bottlenecks. Standard forward-time simulation algorithms like the Gillespie method are inefficient for the investigation of such phenomena due to catastrophic rejection rates. Advanced rare-event techniques like splitting methods and transition-path sampling often suffer from kinetic trapping, path degeneracy, genealogical correlations, or critical slowing down when applied to complex heterogeneous networks.
We propose to overcome this challenge by establishing a novel technique called conditional-path Monte Carlo (CPMC), inspired by loop algorithms from equilibrium condensed-matter physics. By employing non-local updates on spacetime clusters without rejections, CPMC generates a Markov chain of trajectories that all strictly respect the targeted macroscopic boundary conditions like the occurrence of a massive network failure. We demonstrate the framework's potential by performing a simple risk factor analysis for rare large-scale epidemic outbreaks in SIS dynamics on kinship networks.
\end{abstract}

\maketitle

\section{Introduction}
Stochastic dynamics in complex networks and many-body systems play a central role in diverse scientific disciplines. In many real-world systems, the most consequential macroscopic events do not reflect average behavior, but rather emerge from a rare confluence of localized stochastic processes. In non-equilibrium statistical mechanics, massive fluctuations drive the escape from metastable states \cite{Haenggi1990-62}, such as nucleation in kinetic spin models \cite{Binder1987-50,Rikvold1994-49} or complex fluids \cite{tenWolde1997-277}, and they dictate the evolution of glassy systems \cite{Merolle2005-102,Garrahan2007-98,Chandler2010-61}.

Beyond pure physics, cascading failures in IT infrastructure or financial systems often originate from highly improbable combinations of local faults \cite{Albert2000-406,Newman2002-66,Gai2010-466,Haldane2011-469,Acemoglu2015-105}. Zoonotic spillovers of potentially dangerous pathogens from animals to humans are frequent \cite{Parrish2008-72,LloydSmith2009-326,Wasik2019-374,Gray2021-13,Sanchez2022-13}, but widespread epidemics only emerge when local disease surges manage to pass bottlenecks in the heterogeneous contact network \cite{PastorSatorras2015-87,LloydSmith2005-438,Salathe2010-6}. Similarly, strong shifts in societal opinions or collective behavior can be driven by a rare alignment of local and global influences \cite{Castellano2009-81,Watts2002-99}.
Conducting risk factor analysis, identifying structural vulnerabilities, and designing intervention strategies for such pivotal events require the investigation of the dynamics conditioned on the rare macroscopic outcomes.

While classical deterministic compartmental models describe coarse-grained expectations by assuming the system to be sufficiently homogeneous and well-mixed \cite{Anderson1991,Keeling2008,Hethcote1989,Hethcote2000-42,Brauer2019}, they generally fare badly at predicting the emergence of rare macroscopic events \cite{Centola2007-113,Buldyrev2010-464,Ioannidis2022-38} because they wash out the system's heterogeneity \cite{Goldenfeld1999-284,Olinky2004-70,Vespignani2011-8,Rock2014-77,Pellis2015-10,Leventhal2015-6} and neglect the stochasticity of the dynamical processes \cite{Durrett1994-46,Goldenfeld1999-284,Keeling2000-203}.

\begin{figure}[t]
    \centering
    \includegraphics[width=\columnwidth]{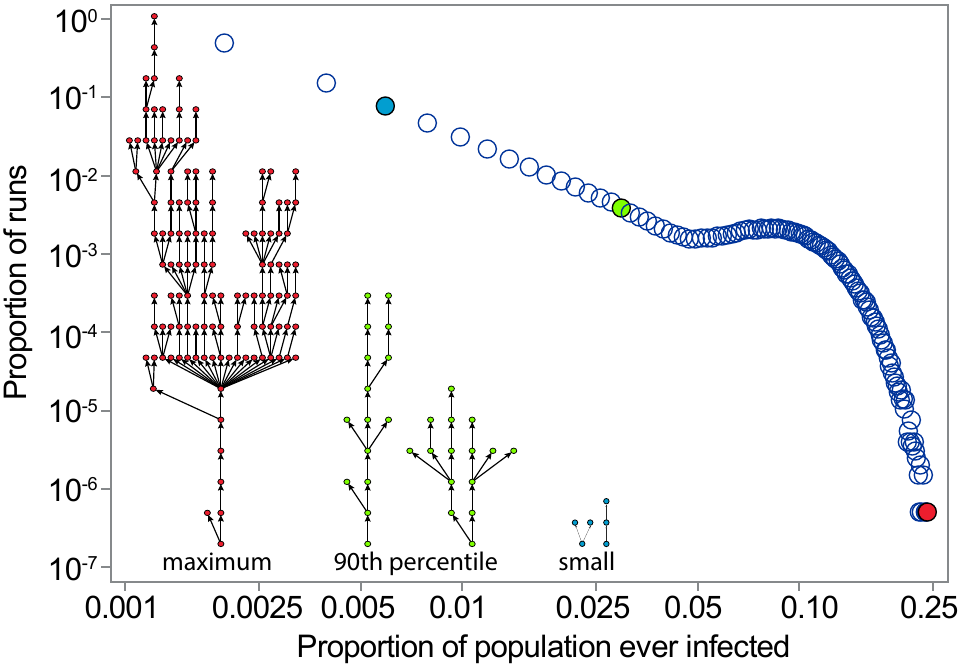}
    \caption{\label{fig:SIRcaseStudy}\textbf{Case study of epidemic outbreak sizes.} For the real-world contact network of a village in rural India with $N=519$ nodes (vertex degree average 9.7 and standard deviation 7.0) \cite{Mohanan2020}, we simulate a susceptible-infectious-recovered (SIR) model with synchronous updates, infection probability 0.015, and a fixed recovery time of 13 rounds (days). The figure shows the distribution of the outbreak sizes and exemplary transmission trees. In the vast majority of cases, the epidemic fades, reaching fewer than five households. In rare cases, however, the transmission proceeds to well-connected nodes which then spread to many, reaching more than 20\% of the households.}
\end{figure}

Network models offer much greater realism \cite{Centola2007-113,Salathe2010-6,Haldane2011-469,Ajelli2010-10,Chowell2016-18}. Traditional individual-based simulation techniques, such as the Gillespie method \cite{Gillespie1977-81} and kinetic Monte Carlo \cite{Bortz1975-17}, are forward-time stochastic simulation algorithms (SSA). While highly efficient for exploring typical, unconstrained dynamics, SSA suffers from catastrophic rejection rates when simulating rare events. The vast majority of generated paths fail to display the targeted events and are discarded. As illustrated in Fig.~\ref{fig:SIRcaseStudy} for epidemic dynamics on a real-world contact network \cite{Mohanan2020}, most disease surges fade quickly. Only a small percentage manages to spread to a macroscopic fraction of the network.

To circumvent the massive rejection rates of forward-time SSA, several valuable rare-event sampling techniques have been developed, yet they face severe challenges for complex network systems \cite{Barthel2026_08}: The weighted stochastic simulation algorithm (wSSA) \cite{Kuwahara2008-129,Gillespie2009-130} relies on importance sampling by artificially biasing local reaction propensities, but often suffers from weight degeneracy and kinetic trapping. Splitting methods, like forward flux sampling (FFS) \cite{Allen2006-124a,Allen2006-124b,Allen2009-21}, repetitive simulation trials after reaching thresholds (RESTART) \cite{VillenAltamirano1991-15,VillenAltamirano1994,VillenAltamirano2002-13}, and weighted ensemble (WE) simulation \cite{Huber1996-70,Rojnuckarin1998-95,Zhang2010-132,Donovan2013-139,Zuckerman2017-46}, branch trial trajectories that cross intermediate thresholds. However, in large heterogeneous networks, the employed scalar order parameters generally mask hidden topological barriers, causing splitting methods to suffer from path degeneracy, critical slowing down, and genealogical correlations \cite{Allen2009-21,Zuckerman2017-46,Aristoff2018-52}. Finally, transition path sampling (TPS) \cite{Dellago1998-108,Dellago1998-108b,Bolhuis2002-53,Dellago2002-1,Bolhuis2021-4} operates directly in the space of trajectories via local ``shooting'' or ``shifting'' moves, but in constrained network systems, these local updates often suffer from critical slowing down as the algorithm struggles to mutate the topological core of the trajectory without violating physical rules.

To overcome the limitations of forward-time SSA, splitting methods, and local path updates, we introduce conditional-path Monte Carlo (CPMC). Inspired by loop algorithms in quantum Monte Carlo \cite{Sandvik1999-59,Evertz2003-52}, CPMC operates on entire trajectories, with specific challenges for trajectory updates arising from causal structures due to the irreversibility of stochastic dynamics. Instead of proposing local, step-by-step changes, CPMC maps the current trajectory to an intermediate graph configuration that decomposes the entire spacetime volume into connected clusters. By employing Swendsen-Wang-like cluster updates \cite{Swendsen1987-58}, the algorithm simultaneously alters the states of the non-local spacetime clusters. This allows CPMC to generate a Markov chain of trajectories that all strictly respect the desired macroscopic constraints, bypassing the massive rejection rates and critical slowing down that plague SSA-based methods and TPS, as well as the path degeneracy problem and genealogical correlations inherent to splitting methods.

\section{Conditional-path Monte Carlo}
\begin{figure*}[t]
\label{fig:E2updateN5}
\centering
\includegraphics[width=\textwidth]{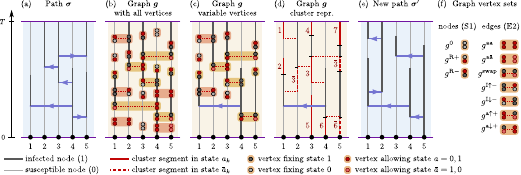}
\caption{\textbf{Example for a cluster update in SIS dynamics.} (a) A trajectory $\vs$ on five nodes with node 4 as patient zero and three infected nodes at the final time. (b) A compatible graph $\vg$. Vertices at state-changing events in $\vs$ are indicated by orange boxes and background vertices are indicated by red boxes. (c) To simplify cluster identification for the human eye, we have removed all vertices that ultimately have no free state. (d) Cluster representation of the graph $\vg$, where each red solid time segment with label ``$k$'' carries the state $a_k=0,1$ of a free cluster $\C_k$ and each red dashed time segment with label ``$\bar{k}$'' carries the negated state $\bar{a}_k=1,0$ of cluster $\C_k$. (e) Flipping the states of all clusters except $\C_5$ with respect to their state in $\vs$, we obtain the new trajectory $\vs'$, where both the trunk and leaves of the infection tree have changed substantially.}
\end{figure*} 
We consider Markovian stochastic dynamics on a network of $N$ nodes, where the system state at time $t$ is given by $\vs^t=(\sigma^t_1,\sigma^t_2,\dotsc,\sigma^t_N)$, and $\sigma_i^t=0,\dotsc,d-1$ are $d$ possible states for node $i$. A continuous-time trajectory spanning $t \in [0,T]$ is denoted by $\vs\in[0,T]\times d^N$. The dynamics are driven by elementary stochastic state changes $\vs_x := (\vs_\vx^{t-}\to\vs_\vx^{t+})$ occurring at spacetime locations $x=(t,\vx)$ with physical event rates $\omega_x(\vs_x)$, where the previous state $\vs_\vx^{t-}$ of some nodes $\vx=(x_1,\dotsc,x_n)$ changes to $\vs_\vx^{t+}$.

As a specific example consider a susceptible-infectious-susceptible (SIS) model \cite{Hethcote1989,Hethcote2000-42,Brauer2017-2,Rock2014-77,Brauer2019} with $d=2$, infection rates $\alpha_{i,j}$ on edges and recovery rates $\gamma_i$. Node $i$ recovering at time $t$ from an infection corresponds to the state change $\vs_x=(\sigma^{t-}_i\to \sigma^{t+}_i)=(1\to 0)$ at location $x=(t,i)$, which happens at rate $\gamma_i$. An infection $\vs_x=(\sigma^{t-}_i,\sigma^{t-}_j\to \sigma^{t+}_i,\sigma^{t+}_j)=(1,0\to 1,1)$ at $x=(t,i,j)$ is associated with the event rate $\alpha_{i,j}$.

When investigating rare events, we restrict the space of trajectories using an indicator function $C(\vs)$ that enforces macroscopic constraints, such as specific initial conditions and outbreak-size thresholds. The unnormalized probability density of a valid trajectory is
\begin{equation}\label{eq:def-P}
P(\vs) = C(\vs) \Big(\prod_{x\in\vs}\omega_x(\vs_x)\Big)e^{-\int_0^T\ud t\,\Lambda^t(\vs^t)},
\end{equation}
where the product runs over all locations $x$ of state-changing events in $\vs$, and $\Lambda^t(\vs^t) = \sum_\vx \Lambda_x(\vs_\vx^t)$ is the total escape rate from state $\vs^t$ with $\Lambda_x(\vs_\vx^t)=\sum_{\tilde{\vs}_x}\delta_{\tilde{\vs}_\vx^{t-},\vs_\vx^t}\,\omega_x(\tilde{\vs}_x)$, where the sum runs over all possible state changes at location $x$.
Simulating Eq.~\eqref{eq:def-P} directly via forward-time SSA results in catastrophic rejection rates because the vast majority of generated paths fail to satisfy $C(\vs)$.

To efficiently sample from $P(\vs)$, CPMC abandons step-by-step local integration. Instead, it generates a Markov chain
\begin{equation}\label{eq:pathChain}
	\vs[1]\xrightarrow{\W} \vs[2]\xrightarrow{\W} \vs[3]\xrightarrow{\W} \dotsc
\end{equation}
of trajectories that all strictly respect $C(\vs)$, operating via non-local spatiotemporal cluster updates. The trajectory update passes through an intermediate graph configuration $\vg$ in two probabilistic steps
\begin{equation}
	\vs \xrightarrow{(a)} \vg \xrightarrow{(b)} \vs'.
\end{equation}
This process is illustrated in Fig.~\ref{fig:E2updateN5} for a simple network.

The graph $\vg$ decomposes the entire spacetime volume $[0,T]\times\{1,\dotsc,N\}$ into connected clusters $\{\C_k\}$, where each cluster possesses a base state $a_k$ that maps to physical node states via an injective function $\sigma_i^t=\sigma_i^t(a_k)$. Clusters with $d_k \geq 2$ allowed states $a_k$ are termed \emph{free} clusters.

In step (a), a graph $\vg$ is constructed by assigning graph vertices $\{g_x\}$ to spacetime locations $x$, where state changes do occur in $\vs$ or could occur. For each state-changing event $\vs_x$ in $\vs$, the algorithm assigns a ``physical vertex'' $g_x$ with probability $\propto\nu_x(g_x)\Delta(\vs_x,g_x)$, where $\nu_x(g_x)$ is a vertex insertion rate similar to the event rates $\omega_x(\vs_x)$, and $\Delta(\vs_x,g_x) \in \{0,1\}$ indicates compatibility between the vertex and the local state dynamics $\vs_x$. With rates $\nu_x(g_x)\Delta(\vs_x^{\const},g_x)$, the algorithm then inserts further ``background vertices'' via a Poisson point process \cite{Daley2003} on time intervals where the node states remain constant, corresponding to null-events $\vs_x^{\const}=(\vs_\vx^t\to \vs_\vx^t)$. Each vertex enforces local bijective state relations or restricts the allowed states of the involved nodes. For example, an infection event $(1,0\to 1,1)$ is compatible with the vertex $(1,0\to 1,a)$, where changing the cluster variable $a$ from 1 to 0 corresponds to removing the infection event. The infection event is also compatible with the vertex $(a,\bar{a}\to 1,1)$, where $\bar{a}$ is the negation of $a$ and changing $a$ from 1 to 0 corresponds to swapping the source node of the infection.

The mapping $\vs\to\vg$ is governed by a joint path-graph weight
\begin{equation}\label{eq:def-J}
	J(\vs,\vg)=C(\vs) \Big(\prod_{x\in\vg}\nu_x(g_x)\Delta(\vs_x,g_x) \Big)e^{-\int_0^T\ud t\,\Gamma^t},
\end{equation}
where the product runs over all vertex locations $x$ of the graph $\vg$, and $\Gamma^t = \sum_\vx \Gamma_x$ is a state-independent total ``uniformization rate'', introduced to achieve detailed balance.
To guarantee that the path probability $P(\vs)$ is recovered when we marginalize $J(\vs,\vg)$ over all graphs $\vg$ that are compatible with $\vs$, we strictly enforce the transition and uniformization sum rules
\begin{subequations}\label{eq:sumRules}
\begin{align}
	\omega_x(\vs_x) &= \sum_{g_x} \nu_x(g_x) \Delta(\vs_x, g_x)\quad\text{and} \label{eq:sum-trans} \\
	\Gamma_x - \Lambda_x(\vs_\vx^t) &= \sum_{g_x} \nu_x(g_x)\Delta(\vs_x^{\const}, g_x). \label{eq:sum-uni}
\end{align}
\end{subequations}
The described procedure samples $\vg$ according to the transition probability density $W(\vg|\vs)=J(\vs,\vg)/P(\vs)$.

In step (b), CPMC translates the generated graph $\vg$ into spacetime clusters $\{\C_k\}$. Because node states are strictly constant between graph vertices, the trajectory is divided into discrete segments. A union-find (disjoint-set) algorithm iterates through the graph vertices $g_x \in \vg$, binding segments together and determining the state restrictions imposed by the vertices.
For each identified free cluster $\C_k$, the algorithm randomly selects the new state $a_k$ uniformly from the allowed set. Because this state change is executed simultaneously across the entire non-local spacetime cluster, CPMC naturally generates valid mutated trajectories $\vs'$ without suffering from the critical slowing down that can paralyze local path-updating schemes.
This procedure samples $\vs'$ according to the probability density $W(\vs'|\vg)=J(\vs',\vg)/\sum_{\vs\in\vg}J(\vs,\vg)$, which implies that the resulting path-path transition probability density $\W(\vs'|\vs)$ in Eq.~\eqref{eq:pathChain} obeys strict detailed balance
\begin{equation}\label{eq:detailedBalance}
	\W(\vs'|\vs)P(\vs)=\W(\vs|\vs')P(\vs').
\end{equation}

Applying CPMC to specific network dynamics, such as the SIS model, requires careful construction of the graph vertex set, including the determination of vertex insertion rates $\nu_x(g_x)$ and uniformization rates $\Gamma_x$ that solve the sum rules \eqref{eq:sumRules}. An elementary infection event $(1,0 \to 1,1)$ can be matched by various edge vertices. We find that naive vertex choices can lead to strong autocorrelations in the Markov chain \eqref{eq:pathChain}. For example, vertices that enforce equality across time can trigger upstream or downstream ``lock avalanches'' which heavily reduce the number of free clusters, limiting the structural mobility of the trajectory.
A method to avoid this is to employ vertices that decouple past and future. For example, the infection vertex $(1,0\to 1,a)$ insulates the future of the target node from the past, and the source swapping vertex $(a,\bar{a}\to 1,1)$ insulates the past of the two nodes from the future. Suitable vertex sets ensure ergodicity and achieve high structural mobility, even in networks with highly asymmetric transmission rates.
\begin{figure*}[t]
    \centering
    \includegraphics[width=\textwidth]{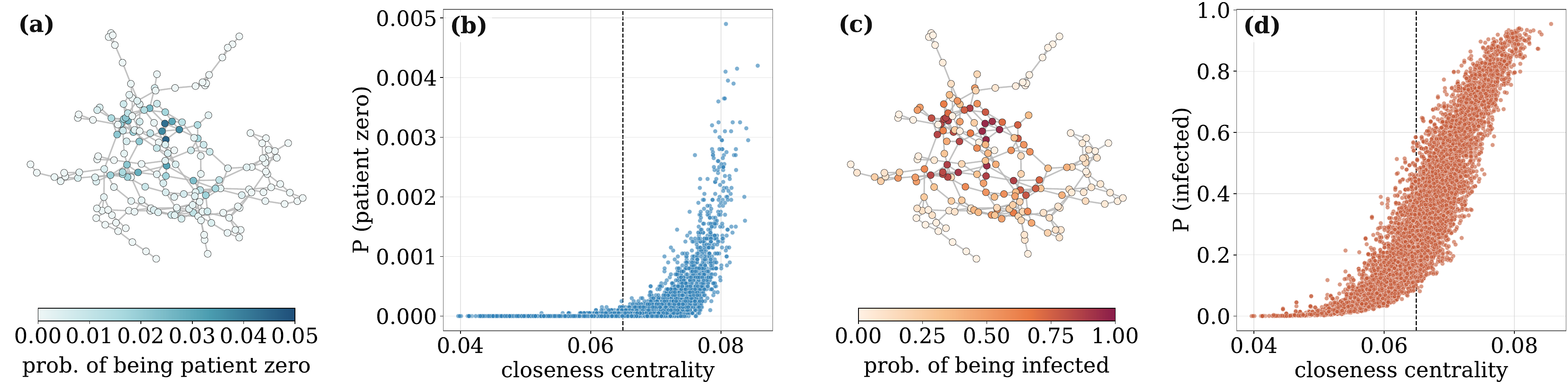}
    \caption{\label{fig:centrality}\textbf{Risk factor analysis for rare SIS outbreaks.} CPMC samples SIS trajectories conditioned on a single infected node at $t=0$ and outbreak size $N^\tI(T)\geq N^\tI_{\min}$, where $N^\tI(t)$ is the number of infected nodes.
    (a) The 185-node kinship network A, with nodes colored according to the probability $p_i^{0}$ [Eq.~\eqref{eq:p0}] of being patient zero.
    (b) $p_i^{0}$ versus the closeness centrality $c_i$ [Eq.~\eqref{eq:centrality}] for the 8623-node network B, each point representing one node.
    (c) Network A colored according to the probability $p_i^\tI$ [Eq.~\eqref{eq:pI}] for getting infected during the outbreak.
    (d) $p^\tI_i$ versus $c_i$ for network B.
    The model parameters are $\gamma=1$ and $(\alpha,T,N^\tI_{\min})=(0.35, 10, 30)$ for network A [panels (a) and (c)] and $(0.6, 20, 1200)$ for network B [panels (b) and (d)]. All results are averaged over $10^5$ trajectories.}
\end{figure*}

An essential advantage of CPMC is its ability to handle constraints $C(\vs)$ without path rejection. Without constraints, the states $a_k$ of all free clusters would be chosen independently at random. With nontrivial constraints, an elegant approach respecting exact detailed balance \eqref{eq:detailedBalance} is dynamic programming: Consider for example an SIS simulation requiring exactly one patient zero at $t=0$ and a large outbreak threshold of at least $N^\tI_{\min}$ infected nodes at time $t=T$. Let $\C_1,\dotsc,\C_K$ denote all free clusters that affect $C(\vs)$ because they intersect with time $t=0$ or $t=T$. Similar to the Knapsack/subset-sum problems \cite{Martello1990,Kellerer2004}, we can build a dynamic programming table $Z$ in a forward pass, where $Z_k(m,n)$ counts the number of valid state combinations $(a_1,\dotsc,a_k)$ that result in exactly $m$ infected nodes at $t=0$ and $n$ infected nodes at $t=T$ in the first $k$ of the $K$ relevant clusters. Because the initial condition demands exactly one patient zero ($m \leq 1$), the table can be aggressively truncated, reducing the computational complexity to $\O(K N_\text{free})$, where $N_\text{free}$ is the total number of nodes at $t=T$ in the $K$ relevant free clusters. In a single backward pass, we then choose cluster states in the order $a_K, a_{K-1}, \dotsc, a_1$, sampling each state based on the marginal probability that it leads to a valid final trajectory with one patient zero and the required outbreak size.

\section{SIS outbreaks in kinship networks}
We apply CPMC to simulate rare large outbreaks in the continuous-time SIS model on synthetic single-gender kinship networks with child reassignment probability $\rho=1$ according to the definition in Ref.~\cite{Zanette2019-9}, also corresponding to the one-parent model of Ref.~\cite{Derrida1991-53}. Specifically, network A comprises $N=185$ nodes with average vertex degree $\langle k\rangle=3.2$ and standard deviation $D(k)=1.5$. Network B has $N=8623$, $\langle k \rangle = 3.3$, and $D(k)=1.6$. We choose homogeneous recovery and infection rates $\gamma_i=1$ and $\alpha_{i,j}=\alpha$ for all $i,j$, a single patient zero, and an outbreak threshold $N^\tI_{\min}$ such that
\begin{equation*}
    N^\tI(0)=1,\ \ N^\tI(T)\geq N^\tI_{\min},\ \ \text{where}\ \
    N^\tI(t)=\sum_i\sigma_i(t)
\end{equation*}
is the number of infected nodes at time $t$. CPMC respects this constraint $C(\vs)$ during every cluster update without rejection.

To determine which nodes carry the highest risk for initiating a large rare outbreak and which nodes are most vulnerable, for each node $i$, we measure the conditional probabilities
\begin{subequations}
\begin{align}
	p_i^0&:=\Pr\big(\sigma_i(0)=1 \mid C(\vs)\big) \label{eq:p0}\\
    p_i^\tI&:=\textstyle\Pr\big(\int_0^T\ud t\,\sigma_i(t)>0 \mid C(\vs)\big) \label{eq:pI},
\end{align}
\end{subequations}
for being patient zero and for getting infected during the outbreak, respectively.
\begin{figure}[t]
    \centering
    \includegraphics[width=\columnwidth]{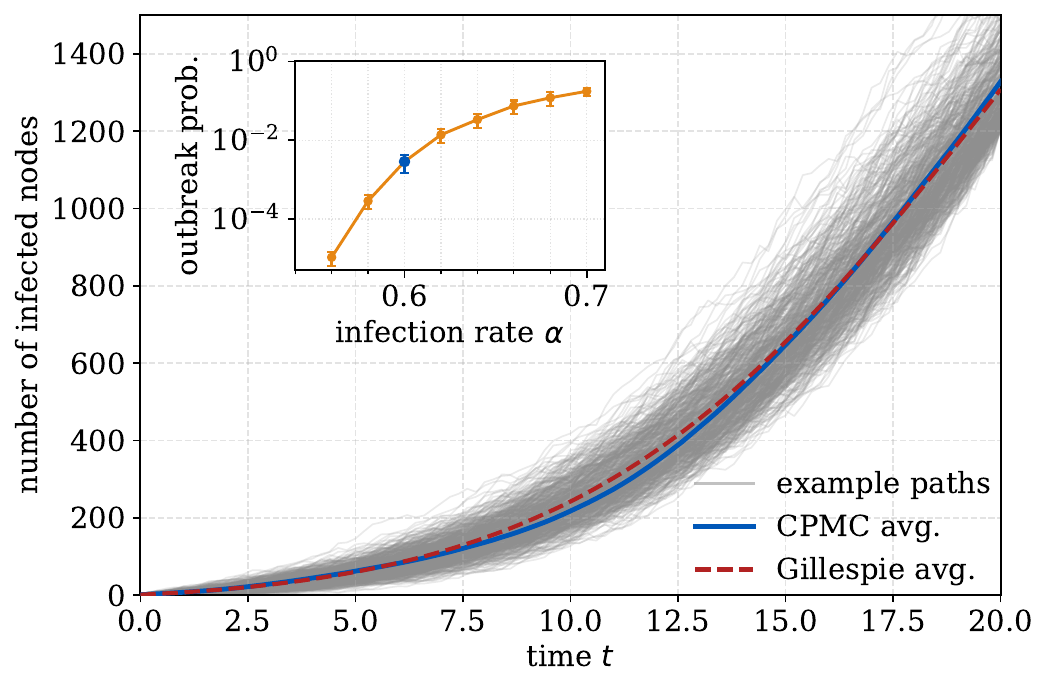}
    \caption{\label{fig:NI}\textbf{Validation of CPMC against Gillespie simulations on the 8623-node kinship network.} Gray curves show representative trajectories satisfying the outbreak condition $N^\tI(T)\geq N^\tI_{\min}=1200$. The mean $\langle N^\tI(t)\rangle$ obtained from CPMC (solid blue) agrees well with the SSA Gillespie simulations (dashed red). Here $\gamma=1$, $\alpha=0.6$, and $T=20$. Inset: unconditioned outbreak probability $\Pr\big(N^\tI(T)\geq N^\tI_{\min}\mid N^\tI(0)=1\big)$ as a function of the infection rate. The blue point marks the rate used in the main panel. Error bars indicate one standard deviation, and results are averaged over $10^5$ samples.}
\end{figure}

For the small network A, Figs.~\ref{fig:centrality}a and \ref{fig:centrality}c indicate visually that, while both distributions are highly heterogeneous, large outbreaks originate preferentially from a few nodes in the network's central region and that the vulnerable group is considerably broader.
To investigate this further, we simulate the SIS model on the considerably larger network B with $(\alpha,T,N^\tI_{\min})=(0.6, 20, 1200)$, examining the relation of $p_i^0$ and $p_i^\tI$ to structural network metrics. Specifically, Figs.~\ref{fig:centrality}b and \ref{fig:centrality}d plot $p_i^0$ and $p_i^\tI$ for all 8623 nodes against their closeness centrality \cite{Bavelas1950-22,Beauchamp1965-10,Newman2010}
\begin{equation}\label{eq:centrality}
	\textstyle c_i=(N-1)/\sum_{j\neq i} \dist(i,j),
\end{equation}
where $\dist(i,j)$ is the length of the shortest path between nodes $i$ and $j$.
The patient zero probability $p_i^0$ starts to rise sharply around 110\% of the average centrality $\langle c_i\rangle=0.065$. In comparison, $p_i^\tI$ shows a broader distribution, starting to rise around $0.75\times\langle c_i\rangle$ and reaching $p_i^\tI=0.5$ around $1.1\times\langle c_i\rangle$. It is of course intuitive that more central nodes have higher $p_i^\tI$ and that large outbreaks are initiated by exceptionally centralized nodes. However, detailed knowledge about relations between structural network metrics and the participation of a node in large outbreaks can play a central role in risk factor analysis, mitigation schemes, and resource allocation.

To validate CPMC, we compare the SIS simulation on the large network B to SSA Gillespie simulations \cite{Gillespie1977-81}. Operating in a regime with relatively high infection rate $\alpha=0.6$, the Gillespie rejection rate (discarding trajectories with $N^\tI(T)<N^\tI_{\min}$) is high but still manageable. Figure~\ref{fig:NI} compares the conditional mean infection counts $\langle N^\tI(t) \rangle_{C(\vs)}$. The data agree well over the entire time interval.

We attribute the remaining small deviations around $t=11$ and $t=20$ to (a) the limited number of samples, (b) upstream lock avalanches caused by the employed edge vertex set \textsc{E1} from Ref.~\cite{Barthel2026_08}, which result in larger autocorrelations in the Markov chain \eqref{eq:pathChain}, and (c) enforcing the boundary constraints through conditional state-locks at $t=0$ and $T$ that slightly break detailed balance.
These issues and ways to drastically improve them by switching to the alternative edge vertex \textsc{E2} and the dynamic programming scheme for constraint satisfaction are discussed in Ref.~\cite{Barthel2026_08}. Here, autocorrelations were mitigated by a parallel tempering scheme \footnote{J.\ Sun and T.\ Barthel, in preparation}. Comparisons of CPMC with different vertex sets and associated vertex insertion rates $\nu_x(g_x)$ will be presented in future work.

The inset of Fig.~\ref{fig:NI} shows the outbreak probability as a function of the infection rate $\alpha$. At $\alpha=0.6$, the targeted large outbreaks occur with a probability of order $10^{-3}$, implying hundreds of discarded Gillespie samples for each accepted outbreak trajectory. The probability drops exponentially with decreasing $\alpha$. In contrast, every single CPMC sample satisfies the prescribed constraint. CPMC provides direct access to rare-event trajectories without the vanishing acceptance rate of traditional forward-time SSA.

\section{Conclusion}
We have described conditional-path Monte Carlo (CPMC), a highly efficient rare-event sampling framework for stochastic dynamics on complex networks. It applies rejection-free non-local updates for spacetime clusters of the entire trajectory and dynamic programming to build a Markov chain of trajectories, each respecting the macroscopic constraints of the targeted class of events. In this way, CPMC circumvents the vulnerabilities of standard forward-time algorithms and traditional rare-event techniques when faced with large heterogeneous networks, rigid topological bottlenecks, or complex boundary conditions.

As demonstrated for the SIS model on kinship networks, CPMC promises to further our understanding of important rare phenomena in physics, biology, engineering, social science, and finance, and to enable an efficient identification of critical risk factors for catastrophic events. Uncovering such risk factors and early indicators can lead to better prevention and control strategies.

Further details and derivations for the CPMC method as well as comparisons to exact solutions for dynamics on small networks are presented in the technical companion paper \cite{Barthel2026_08}.

There are several routes for future methodological developments: It is fairly straightforward to adapt the SIS graph vertex sets to other models; some care is necessary in the design to avoid upstream or downstream lock avalanches. Parallelization strategies to further increase computational efficiency are described in Ref.~\cite{Barthel2026_08}. While the presented framework already covers temporal networks \cite{Holme2012-519} and systems with time-dependent event rates, details for an efficient implementation need to be worked out. To minimize autocorrelation times and push the boundaries of rare-event simulation in massive real-world networks, future work will explore combining CPMC with parallel tempering \cite{Note1} and adaptive machine learning algorithms to optimize vertex rates during the burn-in phase of the trajectory Markov chain.

\section{Acknowledgments}
We gratefully acknowledge discussions with Caterina de Bacco, Hana Jang, Jianfeng Lu, Charles Nunn, Joshua Socolar, and participants of the international workshop ``Quantitative Methods for Dynamics on Networks'' 2024 in Los Alamos (organized by the CNLS at Los Alamos National Laboratory) as well as support by the U.S.\ National Science Foundation through grant no.\ DMS-2344576 and by the Duke Population Research Center (DPRC) through the U.S.\ NICHD grant no.\ P2C-HD0065563.

\end{document}